# Integrated laser heating stage with active geometry modulation for simultaneous in-situ X-ray transmission and evolved gas analysis of molten liquids

Chao Ang[1], Zhenqi Jiao[1], Yue Yang[1], Na Yu[1], Xuerong Liu[1] Yi Hu[2], Zhu-Jun Wang[1]

1. School of Physical Science and Technology, ShanghaiTech University, Shanghai 201210, China.
2. Gaoke Shine City Mocha Compound, The 5th Road of Zhangba Avenue, Gaoxin District, Xi'an, Shaanxi 710076, China.

## Abstract

We report the design and development of a compact, integrated laser heating stage tailored for in-situ high-temperature X-ray transmission studies of molten oxides. In horizontal beam geometries, widely used in both laboratory and synchrotron facilities, the natural spreading (wetting) of molten samples on substrates significantly reduces the effective vertical optical path length, detrimental to signal quality in transmission-mode measurements. To overcome this limitation, we introduced a thermocouple-assisted active geometry modulation technique. This method mechanically lifts the spreading melt into a liquid bridge via surface tension, optimizing the transmission path length while simultaneously enabling in-situ temperature monitoring. The device features a triple-fiber coupled laser head with high power density, a precision closed-loop Proportional-Integral-Derivative temperature control system (stability ±0.2°C), and an atmosphere-controlled vacuum chamber coupled with a mass spectrometer. This integration allows for simultaneous evolved gas analysis, enabling the correlation of structural phase transitions with chemical volatilization or reaction dynamics. Validated by tracking the melting kinetics of a multi-component glass precursor, this versatile setup provides a comprehensive solution for high-quality data acquisition in X-ray transmission experiments across various sources.

## Introduction

The structural evolution and physicochemical properties of molten oxides and glass-forming liquids at high temperatures are of fundamental interest in fields ranging from magmatic petrology to materials engineering[1-3]. To probe the atomic-scale structure of these disordered systems under extreme thermal conditions, non-destructive X-ray techniques operating in transmission geometry - such as X-ray diffraction (XRD) and absorption spectroscopy—are indispensable tools[4-6]. However, establishing a stable and compatible sample environment for these measurements remains technically challenging, particularly when adapting to the horizontal beam configurations common in modern laboratory diffractometers and synchrotron endstations.

A critical limitation in such horizontal setups is the wetting behavior of silicate melts. At high temperatures, melts tend to spread extensively on refractory substrates like sapphire or quartz, forming thin films with negligible vertical thickness[7, 8]. This flattened geometry results in an insufficient optical path length for the horizontally incident X-ray beam, drastically reducing the scattering volume and introducing substantial background noise from the substrate interaction[9]. While aerodynamic levitation offers a containerless solution, its bulky instrumentation and gas-dynamics complexity often hinder integration into space-limited X-ray beamlines Similarly[10-12], conventional resistance furnaces suffer from slow heating rates and limited optical access, restricting their utility in dynamic phase transformation studies[13].

To address these issues, laser heating has emerged as a promising alternative due to its rapid heating capability and compact footprint[14]. In this work, we present a compact, integrated laser heating stage specifically engineered to overcome the wetting-induced geometric limitations in X-ray transmission experiments. The core innovation lies in an active sample geometry modulation strategy: utilizing a coated thermocouple probe to mechanically lift the melt, we exploit surface tension to create a "liquid bridge" configuration. This approach effectively optimizes the transmission cross-section for the X-ray beam, regardless of whether the source is a standard laboratory tube or synchrotron radiation. Furthermore, recognizing that high-temperature structural changes are often accompanied by chemical reactions such as decomposition or volatilization, the system is equipped with an atmosphere-controlled vacuum chamber coupled to a quadrupole mass spectrometer (QMS). This integration enables in-situ Evolved Gas Analysis, allowing researchers to simultaneously monitor the release of volatile species and correlate these chemical events with the structural data. Combined with a multi-beam laser source for enhanced power density, this system provides a robust and versatile platform for the holistic characterization of high-temperature melts.

## Result and Discussion

## Design and assembly of the integrated laser heating reaction cell with optimized optical and thermal management

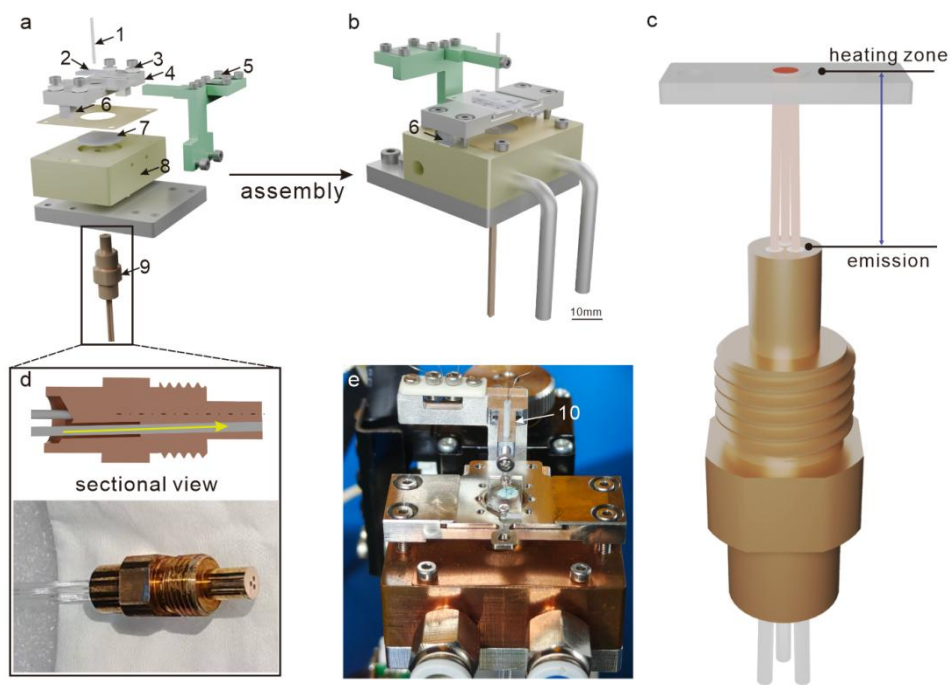

Figure 1 | Design and optical-thermal configuration of the compact laser heating reaction cell.(a) Exploded view showing the modular components of the device. The sample stage (4) is thermally isolated by quartz pillars (6) and ceramic spacers (3). (b) Schematic of the fully assembled device. The reaction cell is built upon a water-cooled copper block (8), with a sapphire window (7) and sapphire carrier (2) supporting the sample. The thermocouple holder (5) aligns the probe (10) via a ceramic tube (1). (c) Schematic of the laser focusing geometry. A triple-fiber array is employed to converge three individual laser beams at the sample heating zone, effectively enhancing the optical power density beyond the limit of a single fiber. (d) Cross-sectional view and photograph of the custom-designed laser head (9). The housing features a stepped cylindrical design with external threads to maximize contact area with the water-cooled block, facilitating efficient conductive cooling to prevent thermal damage to the fiber tips. (e) Photograph of the assembled reaction cell ready for operation.

To enable stable in-situ probing of the microstructure of high-temperature melts, a laser heating reaction cell was designed and developed in this work (Figure 1). The device features a compact, modular layered architecture, primarily consisting of a sample loading system, a multi-beam coupled laser source, and a high-efficiency thermal management base. As shown in Figures 1a and 1b, the device is built upon a high thermal conductivity copper water-cooled block (8) serving as the base. A sapphire window (7) and a sapphire carrier (2) are sequentially assembled above this base; the carrier exploits its excellent high-temperature resistance and optical transparency to directly support the molten sample, while the underlying sapphire window serves to protect the internal optical components. To effectively isolate heat conduction from the high-temperature zone to the external framework, the sample stage (4) is suspended by slender quartz pillars (6) and supplemented with ceramic spacers (3) to construct an effective thermal barrier.

Addressing the power density limitations and thermal damage to fiber tips encountered by conventional single-fiber systems at temperatures exceeding 1000°C, the core laser head (9) features a specialized optical and thermal management design (Figures 1c and 1d). Abandoning the traditional single-beam transmission mode, the laser head adopts a triple-fiber array integration scheme. It utilizes incoherent beam combining technology to precisely overlap three laser beams at a specific convergence angle within the sample zone, thereby significantly enhancing the optical power density at the focal point without increasing the load on individual fibers[15]. Furthermore, to address the issue of heat accumulation at the fiber emission tip during high-power output, the laser head housing is ingeniously designed with a stepped cylindrical structure featuring external threads. This design allows it to be screwed tightly into the water-cooled block, significantly reducing contact thermal resistance by maximizing the metal-to-metal contact area. This facilitates efficient conductive cooling of the fiber tips, thereby ensuring the long-term operational stability of the experimental setup under extreme high-temperature conditions[16].

**Experimental configuration and alignment strategy for in-situ high-temperature X-ray diffraction of molten glass**

Having established the stable heating environment described above, the experiment further addressed the limitations imposed by sample geometry on X-ray detection. In the high-temperature molten state, high-viscosity glassy samples typically exhibit significant wetting behavior on the sapphire carrier surface. As shown in Figure 2a (top), in the absence of external intervention, the melt is governed by the balance between gravity and surface tension, tending to spread across the sapphire surface. This results in an extremely small solid-liquid contact angle ($\theta$)[7]. Such a flattened geometric configuration causes the sample to have a severely insufficient effective thickness or optical path length in the vertical direction (Z-axis) perpendicular to the X-ray incidence. Consequently, when the horizontally incident X-ray beam attempts to penetrate the sample, it is prone to passing directly through the sapphire substrate or grazing the sample surface due to the negligible cross-sectional height, making it impossible to obtain sufficient sample scattering/absorption signals.

To resolve this geometric limitation, a temperature-measuring thermocouple was introduced as an active modulation tool. As illustrated in Figure 2b (top), by leveraging the surface tension and adhesion forces existing between the thermocouple probe and the molten glass, a vertical lifting operation is employed to induce the formation of a meniscus or liquid bridge structure in the melt. This operation effectively counteracts the spreading trend caused by gravity, significantly altering the macroscopic morphology of the melt. It increases the apparent contact angle and vertical exposed volume of the sample on the carrier without changing the material's intrinsic wettability. The resulting conical or neck-like liquid column provides a sufficient projected area for the X-ray beam, ensuring the successful execution of in-situ X-ray characterization under high-temperature molten conditions.

It is worth noting that the thermocouple probe introduced in this setup performs a "dual-functionality" role: "geometric morphology modulation" and "in-situ temperature monitoring". In addition to utilizing the surface tension effect to vertically lift the melt for optimizing the X-ray transmission cross-section, the probe is also responsible for the real-time acquisition of the true temperature of the molten sample.

Given that high-temperature molten glassy samples typically possess strong chemical corrosiveness and reactivity, to prevent interfacial reactions between the metallic thermocouple and the sample during contact measurement, or potential compositional contamination, the surface of the thermocouple probe is pre-coated with a dense alumina ($Al_2O_3$) protective layer. Utilizing the excellent chemical inertness and thermal stability of alumina at high temperatures[17], this coating establishes an effective physical diffusion barrier between the metal matrix and the silicate melt[18]. This design not only guarantees the effectiveness of thermal conduction for accurate temperature measurement but also thoroughly blocks the migration of metallic impurities into the melt, thereby ensuring the compositional purity of the sample and the reliability of the experimental data during in-situ X-ray characterization.

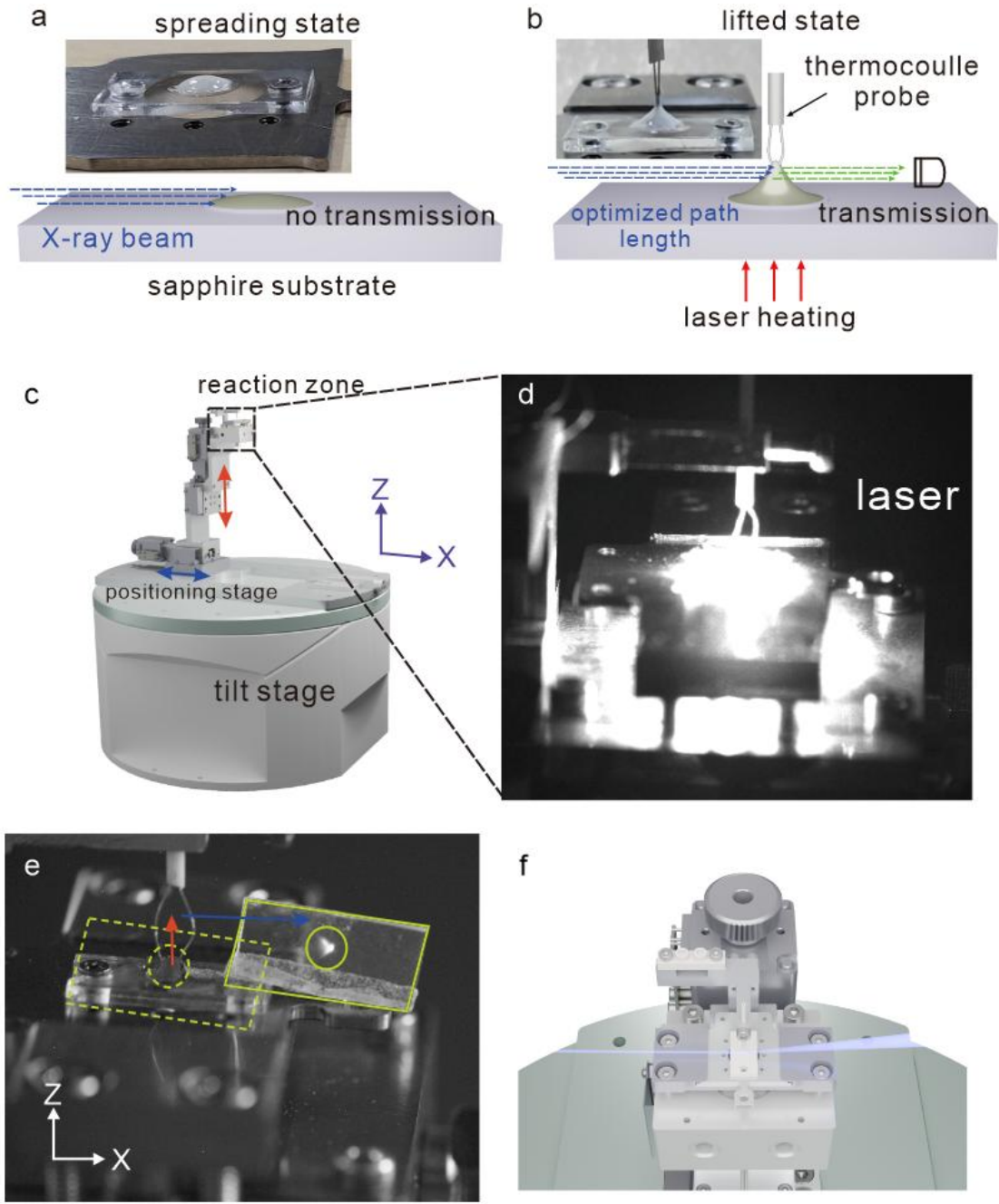

Figure 2 | Sample geometry modulation and precise optical alignment strategy for in-situ X-ray transmission.

(a, b) Comparison of molten sample geometries on the sapphire substrate. (a) In the naturally spreading state, the flat melt profile provides insufficient vertical optical path length, preventing horizontal X-ray transmission. (b) The thermocouple-assisted lifting technique induces a liquid bridge/meniscus, creating an optimized vertical cross-section that allows the X-ray beam to transmit (green arrows) while maintaining laser heating from below. (c) The experimental assembly integrated onto a multi-degree-of-freedom positioning system, comprising a high-precision XZ translation stage mounted on a tilt base for spatial alignment. (d) Close-up view of the laser-heated reaction zone. (e) Visual feedback alignment strategy. The lifted sample is maneuvered via vertical (Z, red arrow) and horizontal (X, blue arrow) translations to precisely align the molten neck with the X-ray beam focus (marked by the green circle), which is pre-calibrated using a YAG scintillator. (f) Schematic of the final experimental geometry, showing the X-ray beam transmitting orthogonally through the aligned, suspended molten sample.

To ensure that the micron-scale (approximately 100 μm) X-ray beam precisely impinges upon the fine neck-like molten region formed by the thermocouple lifting described above, an integrated high-temperature in-situ testing platform was constructed (Figure 2c). The laser heating reaction cell is rigidly mounted onto a high-precision Multi-axis Positioning Stage. This displacement system possesses translation capabilities along the X and Z axes, as well as tilt adjustment, providing three degrees of spatial freedom. This design not only compensates for thermal expansion displacements of the sample at high temperatures but, more crucially, enables the precise maneuvering of the molten sample—specifically the liquid bridge section formed by stretching, which possesses the optimal transmission thickness—onto the Focal Plane of the X-ray beam. This effectively overcomes the spatial matching challenge between the fixed optical path and the dynamic sample.

Given the invisible nature of X-rays and the minute effective detection volume of the sample, a Scintillator-based Visual Feedback Alignment strategy was employed, as illustrated in Figure 2e. First, an X-ray-sensitive Yttrium Aluminum Garnet (YAG:Ce) fluorescent crystal is placed at the sample position. Upon X-ray excitation, it generates a high-intensity visible fluorescence spot[19]. A high-resolution camera coaxial with the setup captures this spot, and its centroid coordinates are marked on the monitor display (indicated as the "Beam Target" by the green circle in the figure).

Subsequently, precise vector translation operations are executed via the positioning stage: first, a vertical elevation along the Z-axis, followed by a lateral movement along the X-axis (as indicated by the red and blue arrows). This precisely maneuvers the melt sample, formed by thermocouple lifting, to the pre-marked beam target position[20]. This "locate-beam-then-move-sample" calibration protocol ensures that the X-ray beam traverses the central region of the sample, characterized by optimized geometry and ideal thickness, thereby maximizing the elimination of geometric errors.

Figure 2f displays the final experimental geometry following calibration. The Horizontally Incident X-ray Beam (high-flux) orthogonally penetrates the glass sample, which is stretched by the thermocouple and maintained in a high-temperature molten non-equilibrium state. Due to the precise alignment achieved via the multi-degree-of-freedom stage, the X-ray beam avoids both the sapphire substrate below and the metallic thermocouple tip above, interacting exclusively with the pristine melt[21]. The resulting Coherent Scattering signals, carrying structural information regarding the atomic arrangement within the melt, are recorded in real-time by a downstream detector array, thereby realizing high-quality high-temperature In-situ XRD analysis.

**Integrated vacuum environmental chamber with dynamic pressure control and in-situ mass spectrometry coupling.**

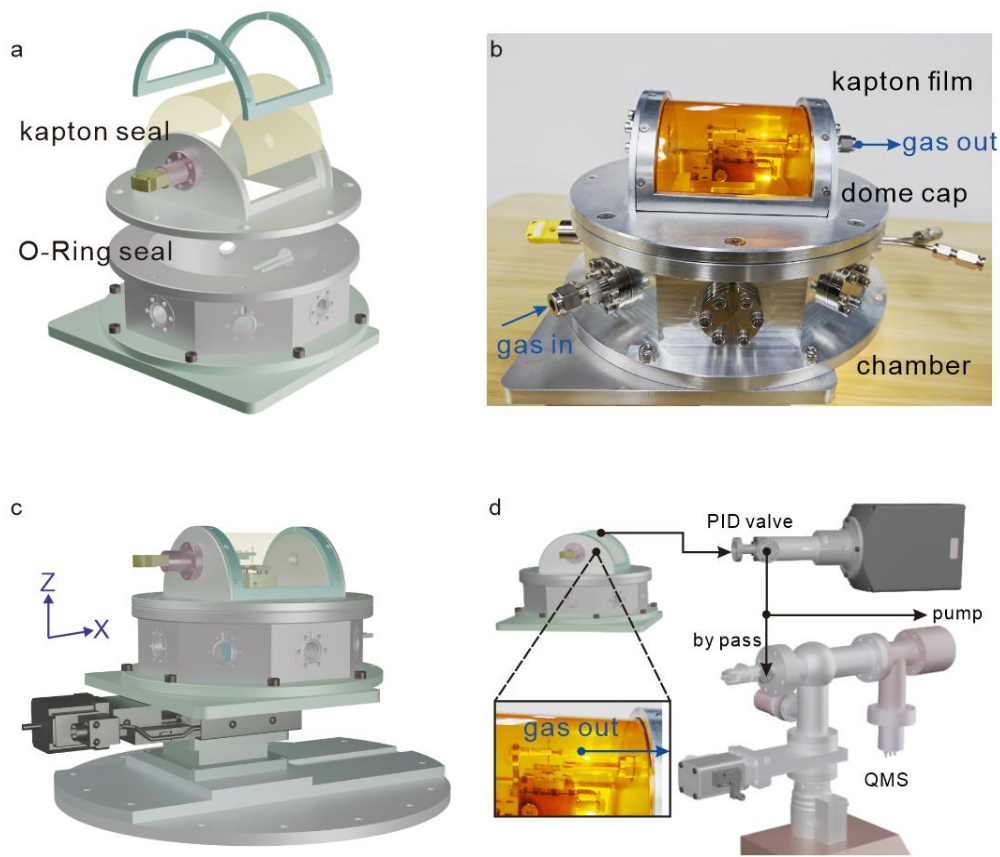

Fig. 3 | Integrated configuration of the atmosphere-controlled vacuum chamber and low-latency gas analysis system.

(a) Exploded view of the chamber components. The panoramic X-ray window is constructed by hermetically sealing a polyimide (Kapton) film onto a hemispherical dome cap. The main chamber is built upon CF16 flange interfaces, integrating modules for vacuum pumping, reactive gas introduction, water cooling circulation, and electrode feedthroughs to meet multi-physics coupling experimental requirements. (b) Photograph of the chamber assembly. Shows the compact external interface layout designed to adapt to the constrained spatial environment of the in-situ stage while optimizing internal gas flow conductance. (c) Multi-dimensional positioning integration system. The chamber is mounted on a high-precision multi-axis positioning stage (featuring X-Z translation and tilt functions) to ensure sub-millimeter precision alignment between the sample micro-zone and the synchrotron X-ray beam. (d) Schematic of dynamic pressure regulation and proximal mass spectrometry sampling interface. The system employs a deep-insertion capillary, extending its tip directly into the sample reaction micro-zone for directional pumping, establishing a tightly coupled "sample-detector" loop. Evolved reaction gases in the main exhaust path are diverted into the Quadrupole Mass Spectrometer (QMS) via a bypass leak valve. This architecture effectively eliminates the dead volume effect of conventional tubing, significantly reducing gas transport lag, thereby enabling high-sensitivity, quasi-real-time monitoring of transient gas-solid reaction products.

To extend high-temperature in-situ research capabilities from open environments to vacuum or specific atmospheric systems, and to effectively isolate the reaction micro-environment from the external space, we developed a dedicated Atmosphere-controlled Chamber System (Fig. 3). This system aims to provide a high-cleanliness experimental background for air-sensitive materials while possessing the capability for high-sensitivity in-situ monitoring of dynamic reaction products[22].

regarding the chamber structural design (Fig. 3a), to minimize X-ray attenuation by the chamber walls and satisfy large-angle scattering detection geometries, the system adopts a Hemispherical Dome Cap configuration. This component features a large-area polyimide (Kapton) film hermetically sealed as a panoramic X-ray window[23, 24]. This geometric design not only provides an expansive, unobstructed field of view but also achieves geometric compatibility with the tilt degrees of freedom of the base: when the bottom stage drives the chamber for optical alignment or Tilt Alignment, the hemispherical profile effectively avoids Beam Occlusion of the incident X-ray beam by the metal skeleton, ensuring optical path clearance and high-throughput transmission over a wide range of angular adjustments[25].

Multiple CF16 standard flange interfaces are precisely arranged around the chamber sidewalls, serving as the hub connecting the internal high-vacuum micro-environment with external control systems. These interfaces fulfill multi-physics field coupling transmission functions: on the material transmission level, supporting precise vacuum pumping and reactive gas introduction; on the energy exchange level, constructing a cooling water circulation loop via integrated fluid feedthroughs to efficiently export waste heat generated by high-power laser heating; on the information transmission level, dedicated electrical feedthroughs establish low-loss signal links, ensuring microvolt-level thermoelectric signals traverse the vacuum barrier without distortion. The entire chamber is rigidly integrated onto a high-precision multi-dimensional motion platform (Fig. 3c), ensuring that the micro-zone sample can still be precisely delivered to the X-ray focal plane under enclosed conditions.

Addressing the dual requirements for atmospheric stability and product analysis timeliness in high-temperature reaction dynamics research, the system constructs a fluid dynamics-based dynamic pressure control and gas analysis loop (Fig. 3d). In terms of pressure control, a PID feedback-controlled throttle valve is connected in series between the vacuum pump and the chamber. By rigorously regulating the effective pumping speed in real-time, a constant pressure environment (Isobaric Condition) is maintained within the chamber under continuous gas flow conditions.

In the critical gas sampling and analysis stage, to overcome the bottleneck of transport lag in traditional tubing, the system adopts a Deep-insertion Capillary design[26, 27]. The tip of the capillary probes directly into the chamber interior, extending to the sample reaction micro-zone for directional pumping, constructing a tightly coupled "sample-detector" loop. Evolved reaction gases in the main exhaust path are diverted into the Quadrupole Mass Spectrometer (QMS) via a Bypass Leak Valve. This architecture effectively eliminates the Dead Volume Effect in the gas path and significantly reduces the transport lag of volatile species. Consequently, mass spectrometry signals can respond with high fidelity to transient chemical changes on the sample surface, successfully achieving precise synchronization of Evolved Gas Analysis and crystal structure detection on a millisecond timescale, providing reliable evidence for elucidating the complex dynamic mechanisms of high-temperature decomposition, oxidation, and volatilization reactions.

## Architecture and performance of the closed-loop PID temperature control system.

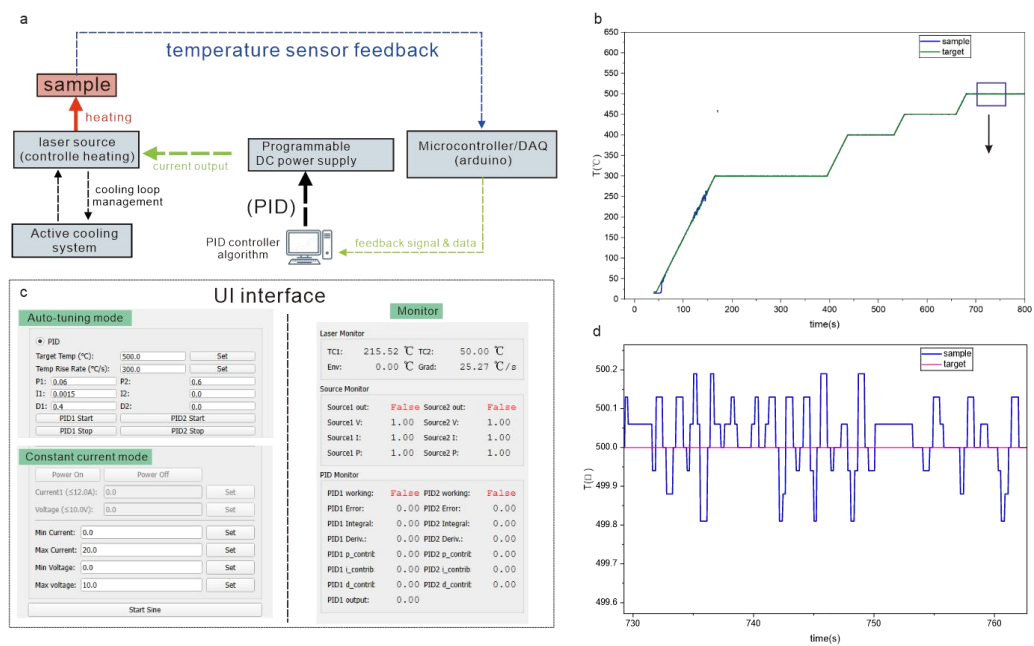

Figure 4 | Implementation of the feedback temperature control system.
(a) Schematic of the hardware control loop. The temperature signal acquired by the thermocouple is digitized by an Arduino-based DAQ and processed by a PC-based PID algorithm to modulate the laser driver current. Safety logic monitors the power-temperature correlation. (b) A representative temperature profile during a heating. (c) Screenshot of the custom-developed user interface (UI) for real-time monitoring and parameter adjustment. (d) The inset highlights the steady-state stability, achieving a precision of ±0.2℃ at high temperatures.

To maximize the hardware performance of the laser heating reaction cell and accommodate the complex thermal environment within the vacuum chamber, a customized Closed-loop Feedback Temperature Control System (Figure 4) was developed[28]. This system aims to resolve the temperature control challenges posed by the non-linear thermal response during high-temperature laser heating, realizing a fully automated process ranging from millisecond-level signal acquisition to precision power regulation[29].

The core of the control loop is built upon a high-precision signal acquisition and drive link. The thermocouple implanted within the sample senses the melt temperature in real-time; its microvolt-level electromotive force signal is first processed by a dedicated digital conversion chip for cold-junction compensation and analog-to-digital conversion, and then transmitted to the host computer (PC) via an Arduino microcontroller over a serial communication port[30]. On the PC side, the integrated control software calculates the deviation between the current measured temperature and the preset target temperature, utilizing a PID algorithm to modulate the output command in real-time. This command directly controls the output current magnitude of a programmable Current Source. Since the laser diodes are in series with the current source and are equipped with independent water-cooled heat sinks to maintain quantum efficiency, minute changes in current are linearly translated into precise adjustments of laser optical power, thereby forming a rapid-response negative feedback regulation loop[31].

Figure 4b displays the User Interface of the system, which not only integrates basic PID parameter adjustment functions but also introduces Adaptive Control Logic. The program allows for the dynamic adjustment of PID coefficients and Ramp Rates based on the experimental load (e.g., sample heat capacity, vacuum level changes) to eliminate Overshoot phenomena. Additionally, the system provides flexible dual-mode operation: Auto-tuning Mode, which strictly follows the set temperature profile for constant temperature control; and Constant Current Mode, which applies a fixed rated current for rapid melting or power testing. To ensure the safety of high-temperature experiments, multiple software and hardware Interlock Mechanisms are embedded in the program. The system sets a hard upper limit on output current to protect the laser, while simultaneously monitoring the correlation between output power and response temperature; if high power output is detected with no corresponding temperature response (implying thermocouple detachment or optical misalignment), the system immediately cuts off the power to prevent overheating accidents[32].

Figure 4c presents a typical heating profile of the system during an actual high-temperature experiment. Benefiting from high-frequency sampling and optimized PID algorithms, the system demonstrates excellent dynamic response capabilities. The magnified inset in the bottom right shows that during the steady-state heating phase at the Set Point, the actual temperature fluctuation of the molten sample is successfully suppressed within a range of 0.2°C. This temperature stability is critical for in-situ XRD experiments requiring long signal acquisition times, ensuring the data precision of lattice parameter measurements and phase transition kinetics analysis.

# Correlating structural phase transitions with chemical decomposition during the synthesis of Pb-Ba silicate

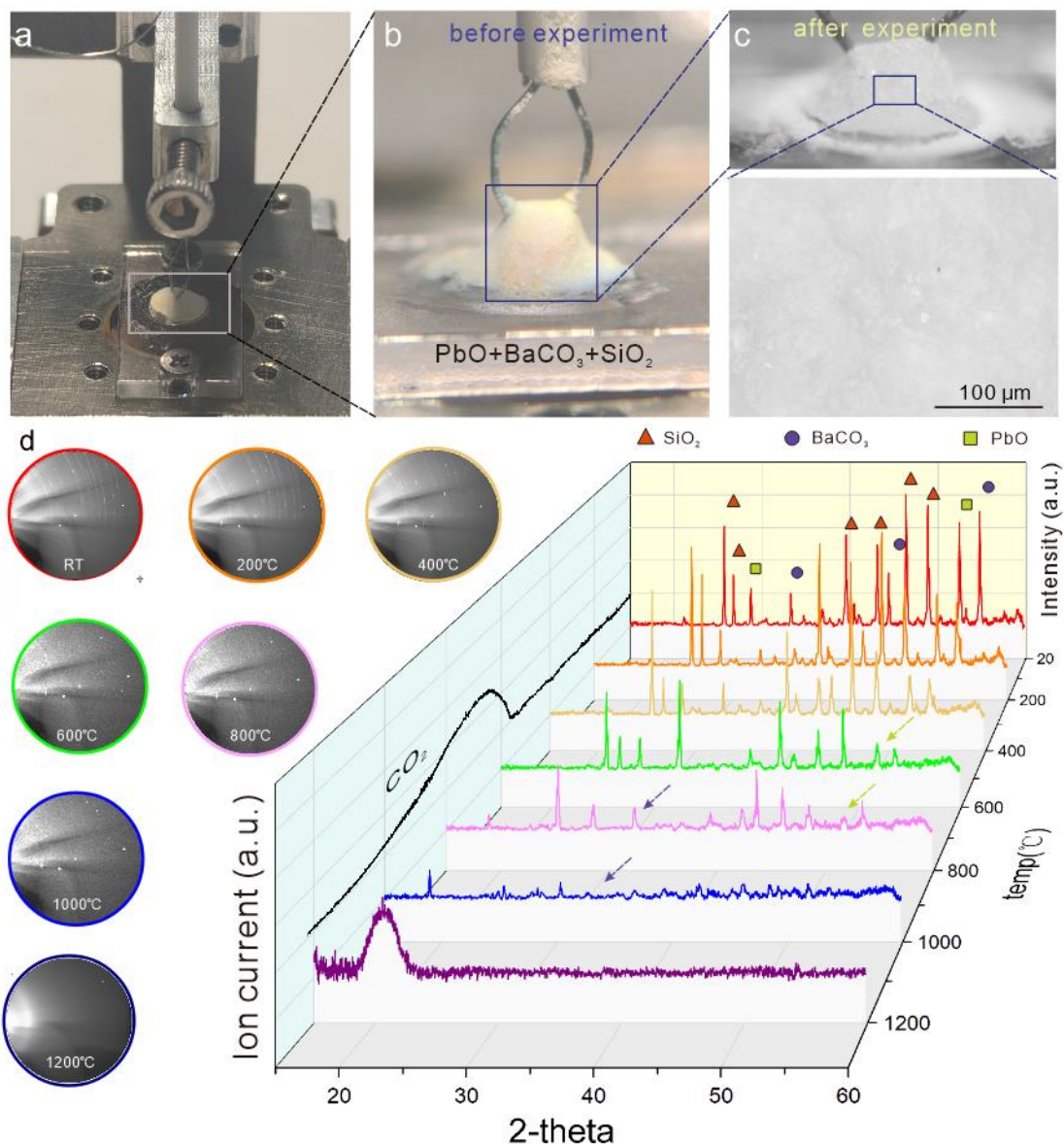

Fig. 5 | Synchronous characterization of in-situ XRD and evolved gas analysis during the Pb-Ba glass firing process.

(a) Sample preparation of the precursor mixture (PbO, BaCO$_3$, SiO$_2$), applied onto the heating carrier via the alcohol dispersion method. (b) Macroscopic appearance of the sample before the experiment, showing the initial powder state. (c) Optical microscopic image of the glass surface after firing; the zoom-in area reveals a homogeneous, amorphous surface texture. (d) Correlation diagram of structural evolution and gas release data during the firing process. The left panel displays the detector images evolving from clear crystalline Debye-Scherrer rings to a broad amorphous diffuse scattering halo with heating. The right panel shows the evolution of XRD diffraction patterns with increasing firing temperature, illustrating the extinction of characteristic diffraction peaks of each crystal phase and the formation of the amorphous envelope. The blue curve projected on the vertical axis is the synchronously collected CO2 evolution mass spectrometry signal, whose peak exhibits high temporal synchronization with the decomposition of the carbonate phase (witherite) and the subsequent formation of the silicate melt.

To further validate the practical efficacy of this integrated laser heating platform in resolving the structural evolution of complex multi-component high-temperature molten oxides, this study selected the PbO–BaCO$_3$–SiO$_2$ ternary system as a typical experimental case to conduct full-process in-situ tracking from multiphase powder to homogeneous glass melt. This system serves as the raw material prototype for ancient Pb-Ba glass (typical high-lead barium glass from the Warring States period to the Han Dynasty in China)[33, 34]. Its multi-step solid-state reactions, carbonate decomposition, and oxide melting behaviors are highly representative, making it a standard benchmark object for studying glass formation and high-temperature phase transition behaviors.

Figure 5 details the structural evolution and gas-solid reaction behavior of the raw material mixture during the programmed temperature increase. At the room temperature stage, the two-dimensional detector image presents clear Debye-Scherrer diffraction rings, corresponding to the three starting crystalline phases (SiO$_2$, BaCO$_3$, PbO) identified in the one-dimensional XRD pattern (right side of Fig. 5d), indicating that the raw materials are in a state of mechanically mixed solids. As the temperature rises, the device clearly captures the stepwise reaction sequence of the system:

In the lower heating zone from room temperature to 600°C, the intensity and position of the characteristic diffraction peaks of each phase show no significant changes, indicating that in this temperature range, the mixed powder remains in a physical stacking state and has not yet induced detectable solid-state reactions or crystal transformations, providing a stable reference state for subsequent high-temperature phase transitions.

In the range of approximately 600-800°C, the intensity of the characteristic diffraction peaks of the PbO phase begins to decrease significantly (marked by the orange arrow in Fig. 5d, right), as it acts as a flux to soften first and participate in the early liquid phase generation reaction. Subsequently, in the temperature window of approximately 600-900°C, the key carbonate decomposition event occurs. The characteristic diffraction peak intensity of BaCO$_3$ decays rapidly and disappears (marked by the blue arrow in Fig. 5d, right); meanwhile, the quadrupole mass spectrometry signal (m/z=44) on the left projection panel shows a synchronous increase in ion current intensity. This high coincidence of structural phase extinction and gas release signals on the time axis confirms in situ that witherite (BaCO$_3$) undergoes a typical thermal decomposition reaction (BaCO$_3$ → BaO + CO$_2$↑) prior to vitrification[35].

When the temperature is further raised to the complete melting zone of about 1100-1200°C, the raw image of the left 2D detector undergoes a decisive transformation: the originally clear and sharp concentric diffraction rings completely disappear, replaced by a broad, diffuse amorphous scattering halo. This change is corroborated in the temperature-dependent XRD diffraction patterns on the right, where all sharp Bragg diffraction peaks, including those of the most refractory SiO$_2$ phase, are completely annihilated and replaced by a broadened and smooth diffuse scattering envelope (Diffuse Scattering Halo)[36, 37].

This characteristic "amorphous hump" is the typical spectral fingerprint of liquid structures with short-range order and long-range disorder, marking that the lead-barium silicate system has completed the multi-step solid-state reactions, carbonate decarbonization, and oxide interdiffusion processes, finally transforming completely into an isotropic, structurally homogeneous liquid glass melt.

The semi-transparent spherical melt obtained at the experimental endpoint (Fig. 5b) and its surface micromorphology (Fig. 5c) further confirm the formation of the vitreous state from both macroscopic and microscopic scales. This case demonstrates that by leveraging the multi-modal real-time synchronization of XRD (structural evolution) and QMS (gas release), the device can precisely resolve the full chain of complex high-temperature reactions covering carbonate decomposition, stepwise melting, amorphization transition, and gas evolution, fully displaying its comprehensive capabilities in the study of high-temperature multiphase reaction kinetics.

**Conclusion and Outlook**

In summary, we have developed an atmosphere-controlled laser heating platform suitable for in situ X-ray transmission experiments. The platform introduces a novel thermocouple-assisted pulling mechanism that effectively resolves the sample wetting issue in horizontal beam path geometries, transforming spreading melts into optimized liquid bridges via active geometry manipulation. This design ensures the sufficient optical path length required for various X-ray techniques, thereby yielding high-quality data. The system features a modular design, incorporating a three-fiber coupled laser head for stable high-power-density heating, while a precise closed-loop PID control system achieves high stability of ±0.2°C at temperatures up to 1200°C. Crucially, the integration of an atmosphere-controlled vacuum chamber with a mass spectrometry interface enables real-time monitoring of evolved gas species and chemical reaction activity.

The open architecture of this experimental apparatus lays a universal foundation for future multi-modal in situ characterization. A key direction for future development is the integration of complementary vibrational spectroscopic techniques, particularly Raman spectroscopy[38] and Fourier Transform Infrared Spectroscopy[39] . While X-ray diffraction (XRD) reveals the average intermediate-to-long-range ordered structure of melts, its combination with Raman or FTIR spectroscopy will allow for the synchronous probing of short-range structural units, chemical bond vibrations, and anionic speciations. This holistic characterization approach will play an important role in acquiring in situ non-equilibrium data and deeply elucidating complex phenomena such as liquid-liquid phase separation, nucleation precursors, and the structural origins of viscosity anomalies in high-temperature oxide melts.

1.   B. Mysen, ISIJ International **61** (12), 2866-2881 (2021).
2.   S. Suzuki, S. Sukenaga, T. Nishi, K. Shinoda and H. Shibata, ISIJ International **63** (5), 767-778 (2023).
3.   B. O. Mysen, Annual Review of Earth and Planetary Sciences **11**, 75 (1983).
4.   J. W. E. Drewitt, C. Sanloup, A. Bytchkov, S. Brassamin and L. Hennet, Physical Review B **87** (22), 224201 (2013).
5.   G. A. Waychunas, G. E. Brown, C. W. Ponader and W. E. Jackson, Nature **332** (6161), 251-253 (1988).
6.   H. Tanida, T. Kobayashi, T. Yaita, M. Kobata, T. Fukuda, A. Itoh, K. Konashi and Y. Arita, Bulletin of the Chemical Society of Japan **98** (2025).
7.   S. Ramamurthy, H. Schmalzried and C. B. Carter, Philosophical Magazine A **80** (11), 2651-2674 (2000).
8.   B. Yu, X. Lv, S. Xiang, C. Bai and J. Yin, ISIJ International **55** (3), 483-490 (2015).
9.   A. Ullrich, K. Garbev and B. Bergfeldt, Minerals **11** (8), 789 (2021).
10.  A. Pack, K. Kremer, N. Albrecht, K. Simon and A. Kronz, Geochemical Transactions **11** (1), 4 (2010).

11. J. K. R. Weber, C. J. Benmore, L. B. Skinner, J. Neuefeind, S. K. Tumber, G. Jennings, L. J. Santodonato, D. Jin, J. Du and J. B. Parise, Journal of Non-Crystalline Solids **383**, 49-51 (2014).

12. C. J. Benmore and J. K. R. Weber, Advances in Science : X **2** (3), 717, 736 (2017).

13. W. J. Campbell, S. Stecura and C. Grain, Advances in X-ray Analysis **5**, 169-190 (1961).

14. Z. Konôpková, W. Morgenroth, R. Husband, N. Giordano, A. Pakhomova, O. Gutowski, M. Wendt, K. Glazyrin, A. Ehnes, J. T. Delitz, A. F. Goncharov, V. B. Prakapenka and H. P. Liermann, J Synchrotron Radiat **28** (Pt 6), 1747-1757 (2021).

15. X. Zhou, Z. Chen, Z. Wang, J. Hou and X. Xu, Optical Engineering **55** (5), 056103 (2016).

16. A. F. El-Sherif, K. Hussein, M. F. Hassan and M. M. Talat, (unpublished).

17. T. Hirata, T. Morimoto, A. Deguchi and N. Uchida, Materials Transactions - MATER TRANS **43**, 2561-2567 (2002).

18. D. Iadicicco, F. Ferre, M. Utili and F. Fonzo, *High performance alumina based gas diffusion barriers*. (2016).

19. T. Martin, A. Koch and M. Nikl, MRS Bulletin **42** (6), 451-457 (2017).

20. D. Carbone and O. Bikondoa, Nuclear Instruments and Methods in Physics Research Section B: Beam Interactions with Materials and Atoms **539**, 127-135 (2023).

21. S. Kobayashi, S. Kawaguchi and H. Yamada, Review of Scientific Instruments **94** (8) (2023).

22. L. Thum, M. Arztmann, I. Zizak, R. Grüneberger, A. Steigert, N. Grimm, D. Wallacher, R. Schlatmann, D. Amkreutz and A. Gili, Review of Scientific Instruments **95** (3) (2024).

23. M. Antimonov, A. Khounsary, S. Weigand, J. Rix, D. Keane, J. Grudzinski, A. Johnson, Z. Zhou and W. Jansma, *Large-area Kapton x-ray windows*. (SPIE, 2015).

24. L. Lurio, N. Mulders, M. Paetkau, P. R. Jemian, S. Narayanan and A. Sandy, Journal of Synchrotron Radiation **14** (6), 527-531 (2007).

25. B. Marco, K. Ivan, G. Alessandro, Y. Yurii and A. Heinz, Journal of Power Sources **477**, 229030 (2020).

26. U. J. Quaade, S. Jensen and O. Hansen, Review of Scientific Instruments **75** (10), 3345-3347 (2004).

27. M. A. Levenstein, C. Chevallard, F. Malloggi, F. Testard and O. Taché, Lab on a Chip **25** (5), 1169-1227 (2025).

28. O. B. Tapar, J. Epp, M. Steinbacher and J. Gibmeier, Metallurgical and Materials Transactions A **52** (4), 1427-1442 (2021).

29. J. Wei and J. Xuchu, Procedia Engineering **43**, 307-311 (2012).

30. R. Anandanatarajan, U. Mangalanathan and U. Gandhi, Experimental Techniques **47** (4), 885-894 (2023).

31. Y. Zhao, Z. Tian, X. Feng, Z. Feng, X. Zhu and Y. Zhou, Sensors **22** (24), 9989 (2022).

32. Y. He, X. Jin, P. Jin, J. Su, F. Li and H. Lu, Photonics **12** (3), 241 (2025).

33. Q. Ma, D. Braekmans, A. Shortland and A. M. Pollard, Journal of Non-Crystalline Solids **551**, 120409 (2021).

34. W. Ha, F. Sun and C. Zhai, PLOS ONE **20** (8), e0331048 (2025).

35. I. Arvanitidis, D. Siche and S. Seetharaman, Metallurgical and Materials Transactions B **27** (3), 409-416 (1996).

36. G. V. P. Bhagath Singh and K. V. L. Subramaniam, Construction and Building Materials **124**, 139-147 (2016).

37. M. C. Rowe and B. J. Brewer, Computers & Geosciences **120**, 21-31 (2018).

38. E. Boccaleri, F. Carniato, G. Croce, D. Viterbo, W. van Beek, H. Emerich and M. Milanesio, Journal of Applied Crystallography **40** (4), 684-693 (2007).
39. Y. Liu, F. Tian, P. Zhou, H. Zhu, J. Zhong, M. Chen, X. Li, Y. Huang, J. Ma and F. Bian, Review of Scientific Instruments **94** (3) (2023).